# Stacking Dependent Optical Conductivity of Bilayer Graphene

Yingying Wang,[†] Zhenhua Ni,[†,‡] Lei Liu,[†,*] Yanhong Liu,[†] Chunxiao Cong,[†] Ting Yu,[†] Xiaojun Wang,[§,⊥] Dezhen Shen,[§] and Zexiang Shen [†,*]

[†]Division of Physics and Applied Physics, School of Physical and Mathematical Sciences, Nanyang Technological University, 21 Nanyang Link, Singapore 637371

[‡]Department of Physics, Southeast University, China 211189

[§]Key Laboratory of Excited State Processes, Changchun Institute of Optics, Fine Mechanics and Physics (CIOMP), Chinese Academy of Sciences, 16 Eastern South Lake Road, Changchun 130033

[⊥]Department of Physics, Georgia Southern University, Statesboro, Georgia 30460

* Corresponding authors, E-mail:

zexiang@ntu.edu.sg

liulei@ntu.edu.sg

# ABSTRACT

The optical conductivities of graphene layers are strongly dependent on their stacking orders. Our first-principle calculations show that while the optical conductivities of single layer graphene (SLG) and bilayer graphene (BLG) with Bernal stacking are almost frequency independent in the visible region, the optical conductivity of twisted bilayer graphene (TBG) is frequency dependent, giving rise to additional absorption features due to the band folding effect. Experimentally, we obtain from contrast spectra the optical conductivity profiles of BLG with different stacking geometries. Some TBG samples show additional features in their conductivity spectra in full agreement with our calculation results, while a few samples give universal conductivity values similar to that of SLG. We propose those variations of optical conductivity spectra of TBG samples originate from the difference between the commensurate and incommensurate stackings. Our results reveal that the optical conductivity measurements of graphene layers indeed provide an efficient way to select graphene films with desirable electronic and optical properties, which would great help the future application of those large scale misoriented graphene films in photonic devices.

Since its first successful isolation from highly oriented pyrolytic graphite in 2004 using micromechanical cleavage method,[1] graphene has attracted more attention because of its fascinating electronic and optical properties, such as its Dirac nature of charge carries,[2-6] its ultra-high electron mobility,[4,7-8] as well as its universal optical or AC (alternating current) conductivity ($G(\omega) = \frac{e^2}{4\hbar}$) value.[9-13]

Many interesting results are obtained from recent optical conductivity studies of graphene layers. *E.g.*, the fine structure constant $\alpha$ ($\alpha = \frac{G(\omega)}{\pi\varepsilon_0 c}$, here, $\varepsilon_0$ is the permittivity of free space and c is the speed of light) can be experimentally obtained from the absorbance measurements of graphene.[11-12] Also, the energetically preferred stacking order, *i.e.* Bernal stacking (ABAB) or orthorhombic stacking (ABCA) of the graphene layers can be differentiated from the calculated conductivity spectrum in the IR range.[14] In addition, the interlayer hopping rate parameter $\gamma_1$ ($\gamma_1 \approx 0.35\,\text{eV}$) of BLG with Bernal stacking can be experimentally obtained from optical conductivity measurements.[15-16]

In this letter, we report a systematic study on the stacking dependent optical conductivities of graphene layers both theoretically and experimentally. The band structure calculations of graphene layers with different stacking sequences are performed using the local-density approximation (LDA) within density-functional theory (DFT), with the Kohn-Sham equations solved with the projected augmented wave method as implemented in the VASP code.[17-19] Here, a kinetic energy cutoff of 400 eV and k-point sampling with 0.05 Å$^{-1}$ separation in the Brillouin zone are used.

Experimentally, the optical conductivity spectra of TBG samples with different stacking geometries (top layer rotates different angle relative to bottom layer) are obtained from their contrast spectra.

Figure 1a gives the calculated electronic band structure of SLG. As shown in this figure, the electronic band of SLG shows the linear dispersion around the Dirac point (K point). The energy span between the conduction and valence bands at M point is ~ 4 eV, slightly lower than the value obtained by angle-resolved photoemission spectroscopy (ARPES) and GW results (4.6 eV), [20-22] as the DFT calculations would normally underestimate the quasiparticle's energy.[21] Figure 1b plots the band structure of BLG with Bernal stacking, where the $\pi$-electron dispersion in the valence and conduction bands splits into two parabolic branches near the Dirac point.[23] Due to the strong interlayer interaction, a band splitting (~ 0.35 eV) appears at the Dirac point.[15] Compared with those from BLG with Bernal stacking, the electronic band structures of TBG samples are much more complicated. Depending on the rotation angle between neighboring layers, TBG samples will be accompanied by the different unit cell and the various interlayer coupling strength, and therefore exhibit abundant electronic properties. Figure 1c shows the band structure of TBG sample with orientation angle of 21.8° (unit cell of 28 atoms). As can be seen, the SLG-like linear electronic dispersion is presented at the Dirac point which agrees very well with other group's results.[24-25] However, the enlarged unit cell of this TBG sample will fold its Brillouin zone, therefore shorten the linear dispersion range as well as reduce the energy span between the conduction band and valence band at M point. As the band gap at M point shifts into the visible light range, *i.e.* ~ 2.77 eV, as shown in Figure 1c, the light absorption behavior of this TBG sample will be altered accordingly.

The optical conductivities of SLG, BLG with Bernal stacking and TBG sample with a 21.8° orientation angle calculated by Kubo formula [10] are shown in Figure 1d which agree very well with other group's results.[10,26] In the visible light range, SLG and BLG exhibit almost universal conductivity behaviors.[10-11,13] On the other hand, the conductivity of TBG sample is frequency dependent which has an additional peak located ~ 2.77 eV. This peak is induced by the electronic transition at the M point. We also calculate the electronic band structures of TBG samples with other orientation angles; top layer rotates 13.2°, 9.4° and 7.3°, individually, relative to the bottom layer. As the orientation angle decreases, more atoms are included in the unit cell and the energy span between conduction and valence states at M point decreases. Their band gaps at M point are 1.81 eV, 1.28 eV, and 0.94 eV for TBG samples with orientation angle of 13.2°, 9.4° and 7.3°, respectively. The electronic transition at M point of these TBG samples will induce an absorption peak in the conductivity spectrum, similar to the result given in Figure 1d. Figure 2a and b give the band structure and density of states (DOS) of TBG sample with orientation angle of 7.3° (unit cell of 244 atoms). As can be seen, the band gap at M point shifts into IR range (0.94 eV). Besides, the band folding and splitting appear in the visible light range resulting in additional features in its DOS spectrum correspondingly as shown in Figure 2b. It is known that the conductivity of graphene is proportional to the joint density of states (JDOS) which can be deduced from its DOS.[27] Therefore, the abundant peaks appearing in the DOS spectrum in the visible light range will induce more peaks in the conductivity spectrum. Based on above discussions, we could know that the stacking sequence (different orientation angle) would typically affect the electronic structures of TBG samples by reducing the band gap at the M point and introducing more bands in the visible light range. Both of them will distinguish the optical conductivities of TBG samples from those of SLG and BLG with Bernal stacking.

To demonstrate the stacking dependent optical properties of graphene layers, Figure 3 plots the absorption behaviour of BLG with Bernal stacking as well as that of TBG sample with orientation angle of $\theta = 21.8^\circ$. We can see that BLG with Bernal stacking shows a constant absorption (~ 4.6%) in the visible light region as a result of universal optical conductivity value.[11] In the ultraviolet light region (300 nm-380 nm), its absorption deviating from constant value is due to the electronic transition near the M point.[26] On the other hand, the TBG sample with 21.8$^\circ$ orientation angle has frequency-dependent absorption behaviour in the visible light range as illustrated in the right side of the figure. Beyond the absorption at the ultraviolet light range, there is an additional absorption peak in the visible light region (highlighted by a red arrow) which is induced by the electronic transition at the M point with transition energy of ~ 2.77 eV. Our calculation results indeed predict TBG sample will present orientation dependent additional absorption peak in the visible light range rather than the constant profile from BLG with Bernal stacking. Since that makes identifying stacking sequence of graphene layers much easy and applicable, therefore, such prediction deserves to be confirmed in careful experiments.

Experimentally, instead of performing absorption measurement, we obtain optical conductivities of graphene layers from their contrast spectra.[13] Compared with the absorption measurement,[11] the contrast measurement can be performed on any substrate; with better spatial resolution of micron scale[28] and do not need suspended samples. The graphene samples are fabricated by micromechanical cleavage and transferred to the SiO$_2$/Si substrate. The TBG samples are prepared by simply flushing de-ionized water across the surface of the substrate, which contains the target SLG.[19,29] In some cases, SLG samples are already folded after cleaving from graphite. The rotation angle between the top layer and the bottom layer can be determined by analyzing the geometry. The crystal axis of SLG can be determined by the graphene edge.[30] Once the angle $\alpha$ between the folded edge and the crystal axis is known, the orientation angle $\theta$, which is the angle of the top layer rotates relative to the bottom layer,[19] can be estimated as $\theta = 2\alpha$ or $\theta = 180° - 2\alpha$.

In the contrast experiments, a tungsten halogen lamp (excitation range from 350 nm to 850 nm, through a 1 mm aperture) is used as the white light source. Both the scattered light and reflected light are collected in the backscattering configuration, using an objective lens with a magnification of 100X and a numerical aperture (NA) of 0.95. The white light spot size is about 1 $\mu m$.[28] The detailed experimental setup can be found in reference [31].

The contrast spectra $C(\lambda)$ can be obtained by [31-34]

$$C(\lambda) = \frac{R_0(\lambda) - R(\lambda)}{R_0(\lambda)}, \tag{1}$$

where $R_0(\lambda)$ is the reflection spectrum from the SiO$_2$/Si substrate and $R(\lambda)$ is the reflection spectrum from the graphene layers on SiO$_2$/Si substrate. According to Fresnel's equation, under normal incidence,

$$R_0(\lambda) = |r_0|^2 = \left| \frac{r_{02} + r_{23} \cdot e^{-2i\phi_2}}{1 + r_{02} \cdot r_{23} \cdot e^{-2i\phi_2}} \right|^2. \tag{2}$$

Here, $r_0$ is the effective reflection coefficient of the air/(SiO$_2$ on Si) interface, and $r_{02} = \frac{n_0 - n_2}{n_0 + n_2}$, $r_{23} = \frac{n_2 - \tilde{n}_3}{n_2 + \tilde{n}_3}$ are the individual reflection coefficients at the air/SiO$_2$ and SiO$_2$/Si interfaces respectively. $\phi_2 = \frac{2\pi \cdot n_2 \cdot d_2}{\lambda}$ is the phase difference when light passes through SiO$_2$ layers. ($n_0$, $n_2$ and $\tilde{n}_3$ are the refractive indices of air, SiO$_2$, and Si, respectively. $d_2$ is the thickness of the SiO$_2$ layer). The optical transfer properties of graphene are governed by the matrix, $M_g = \begin{bmatrix} 1-\beta & -\beta \\ \beta & 1+\beta \end{bmatrix}$,[13,33] and the reflection spectrum from graphene on SiO$_2$/Si as a function of the optical conductivity of graphene, is calculated as:

$$R(\lambda) = \left| \frac{-r_0(1-\beta) + \beta}{-r_0(-\beta) + (1+\beta)} \right|^2 \tag{3}$$

here, $\beta = \frac{c\mu_0 G(\omega)}{2}$ is 188.4 times the optical conductivity; and c and $\mu_0$ are respectively, the speed of light and magnetic permeability in vacuum. Since the contrast spectrum $C(\lambda)$ of graphene layers can be obtained experimentally, the optical conductivity spectrum $G(\omega)$ of graphene layers then can be obtained by solving Eq. (1), (2) and (3).

Figure 4a and b respectively show the optical and atomic force microscopy (AFM) images of pieces of folded graphene layers, which can be taken as TBG samples. Two different folded parts can be seen from the optical image whose boundaries are demarcated by the yellow dashed lines. In Figure 4b, the crystal axis of SLG as well as the folded edge is indicated by white dashed lines. From the geometry analysis, the orientation angle $\theta$ is ~ 22.6º for the left side TBG sample and $\theta$ is ~ 13.7º for the right side TBG sample. Figure 4c shows the contrast spectra of SLG, BLG with Bernal stacking, and TBG samples with $\theta=13.7º$ and $\theta=22.6º$. During the experiments, we have taken contrast spectra from different areas of the TBG sample and the results are the same. For SLG, its contrast peak is located around 557 nm with the peak intensity of ~ 0.1, while that of BLG with Bernal stacking has twice of the SLG's intensity value, which is ~ 0.2.[33] It also can be seen from this figure that the contrast peak of TBG sample has different profile or intensity compared with those of SLG and BLG with Bernal stacking. The optical conductivity spectra of SLG, BLG with Bernal stacking, and TBG samples are then derived from their contrast spectra by solving Eq. (1)-(3).

Figure 5, left column, from bottom to the top, individually shows the conductivities of SLG, BLG with Bernal stacking, and TBG samples with orientation

angle of $\theta=7.5°$, $10.6°$, $12.5°$ in the range of 505 nm (2.46 eV) to 705 nm (1.76 eV), while the right column gives the conductivities of TBG samples with orientation angle of $\theta=13.7°$, $22.6°$, $53.2°$, $54.6°$ and $55.5°$.[35] Here, the normalization of the conductivities has been carried out.[36] As shown in this figure, the conductivity of SLG is frequency independent with a value of $\sim \frac{e^2}{4\hbar}$, while that of BLG with Bernal stacking has twice of the value, *i.e.* $\sim \frac{e^2}{2\hbar}$, consistent with the theoretical predictions.[16] However, the conductivities of some TBG samples behave rather differently from those of SLG and BLG with Bernal stacking. *E.g.*, for TBG sample with $\theta=7.5°$ (close to $\theta=7.3°$ in our calculation), its conductivity is similar as that of SLG in the low energy region (<2.03 eV (610 nm)). Following the increasing of photon energy (>2.25 eV (550 nm)), its conductivity gets higher with the value of $\sim 2.3 \frac{e^2}{4\hbar}$, which is even larger than that of BLG. The observed conductivity of this TBG sample complies with the theoretical prediction very well, whose DOS spectrum given in Figure 2a shows the stronger absorption of the higher energy photons. While for the TBG sample with $\theta=13.7°$ (close to $\theta=13.2°$ in our calculation), there is a broad peak located around 595 nm (2.08 eV), with the height of $\sim 2.52 \frac{e^2}{4\hbar}$. Its conductivity profile also matches quite well with the theoretical result that the electronic transition at M point of TBG sample with $\theta=13.2°$ will induce an absorption peak at ~1.81 eV. Meanwhile, for TBG sample with $\theta=54.6°$, its conductivity shows a sharp increase for the wavelength shorter than 570 nm (>2.18 eV). Those observed orientation-dependent features in the optical conductivity spectra of TBG samples are consistent with their folded electronic bands from the multiple unit cells. Moreover, the conductivities of TBG samples with $\theta$=10.6º, 12.5º, 22.6º, 53.2º and 55.5º, present only frequency independent universal values ($\sim \frac{e^2}{2\hbar}$) in the detection range, similar to that of SLG. The loss of frequency dependent optical conductivity of these TBG samples can be understood by their incommensurate stacking styles. For commensurate stacking of graphene layers, its Moire pattern is periodic, and there exists a unit cell to form its crystal structure.[24] While for those graphene layers with incommensurate stackings, its unit cell is infinite large, *i.e.*, there are no periodic structures. Therefore, incommensurate TBG can be viewed as two individual SLG because of the interaction between the neighbouring layers has been averaged out. As a result, the incommensurate TBG samples will show the optical conductivity with the universal value twice that of SLG. Based on above discussions, TBG samples with the orientation angles of $\theta$ = 7.5º, 13.7º and 54.6º which give additional features in their conductivity spectra can be viewed as commensurate TBG samples, while TBG samples with other orientation angles of $\theta$ = 10.6º, 12.5º, 22.6º, 53.2º and 55.5º whose conductivities are similar as SLG can be viewed as incommensurate TBG samples. We would like to note that we did not perform the DFT calculations on those TBG samples with orientation angles, *e.g.*, 10.6º, 22.6º, 53.2º, 54.6º and 55.5º (as given in our experiments) due to their large unit-cell size or incommensurate stacking style.[24]

Large scale growth of graphene is one of the critical steps towards practical applications of graphene.[1,3-5] Many attempts have been developed to achieve fabricating large scale graphene films such as epitaxial grown on SiC [2,37-38] as well as chemical vapor deposition on Ru[39], Ni[40-41] and Cu.[42] While the stacking order deviating from Bernal stacking is always observed on such large scale graphene films,[24-25,40,43] the desirable frequency independent optical conductivities of those graphene films can't unambiguously be obtained based on our study. However, the contrast measurement, as presented in this work, may distinguish the graphene films with different stackings and that makes the selective use of graphene films with particular electronic and optical properties possible.

In summary, first principle calculations show that the optical conductivities of TBG samples are frequency dependent in the visible light range, contrary to the frequency independent conductivities of SLG and BLG with Bernal stacking. Experimentally, the optical conductivities of TBG samples with different orientation angles are obtained from their contrast spectra. Some TBG samples show additional features in their optical conductivity spectra while others present frequency independent values in the whole detection range. Such controversy of the optical conductivities of TBG samples has been explained by the difference between the commensurate and incommensurate stacking styles. By performing optical conductivity measurements, graphene films with different stacking sequences can be clearly distinguished and selected.

Figure 1

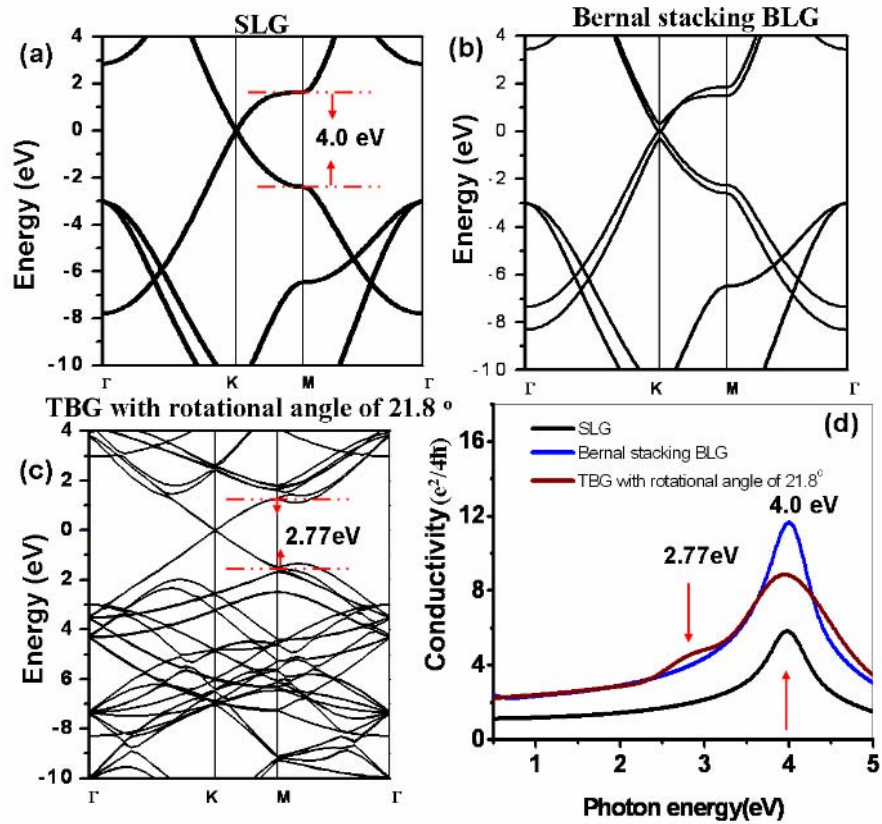

Figure 1. (a) The band structure of SLG. (b) The band structure of BLG with Bernal stacking. (c) The band structure of TBG with orientation angle of $\theta = 21.8^\circ$ (unit cell of 28 atoms). (d) The conductivities of SLG, BLG with Bernal stacking and TBG with orientation angle of $\theta = 21.8^\circ$ calculated by the Kubo formula.

Figure 2

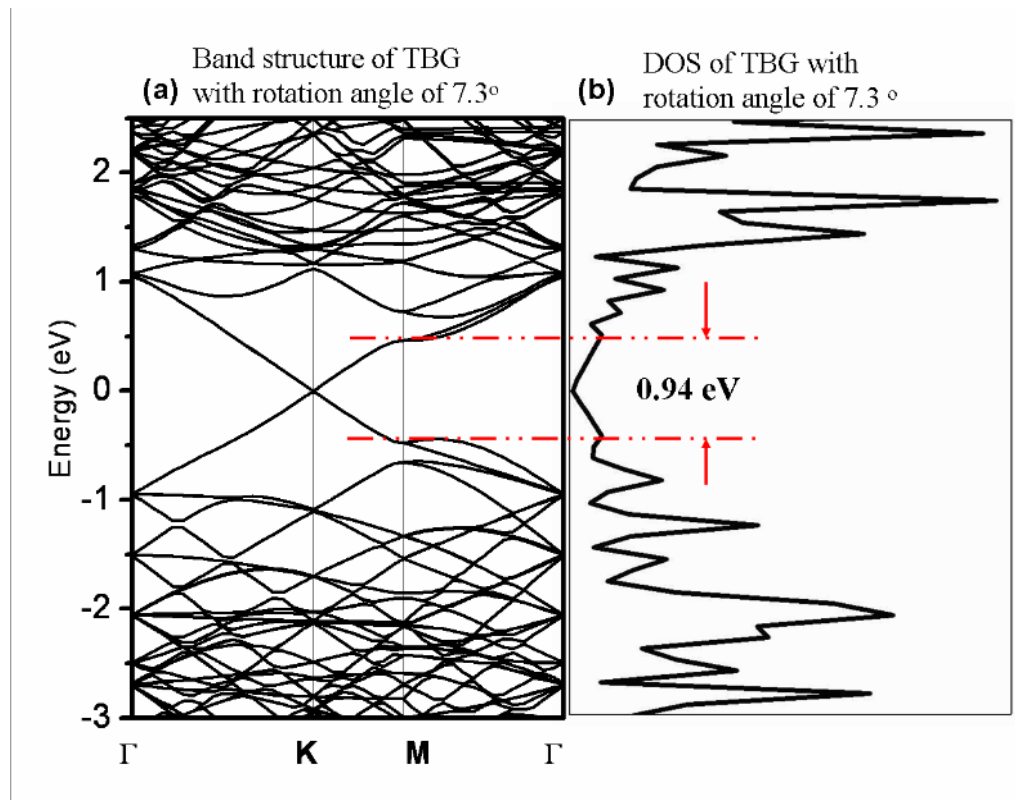

Figure 2. (a) The band structure of TBG with orientation angle of $\theta = 7.3^o$. (b) The DOS of TBG with orientation angle of $\theta = 7.3^o$ (unit cell of 244 atoms).

Figure 3

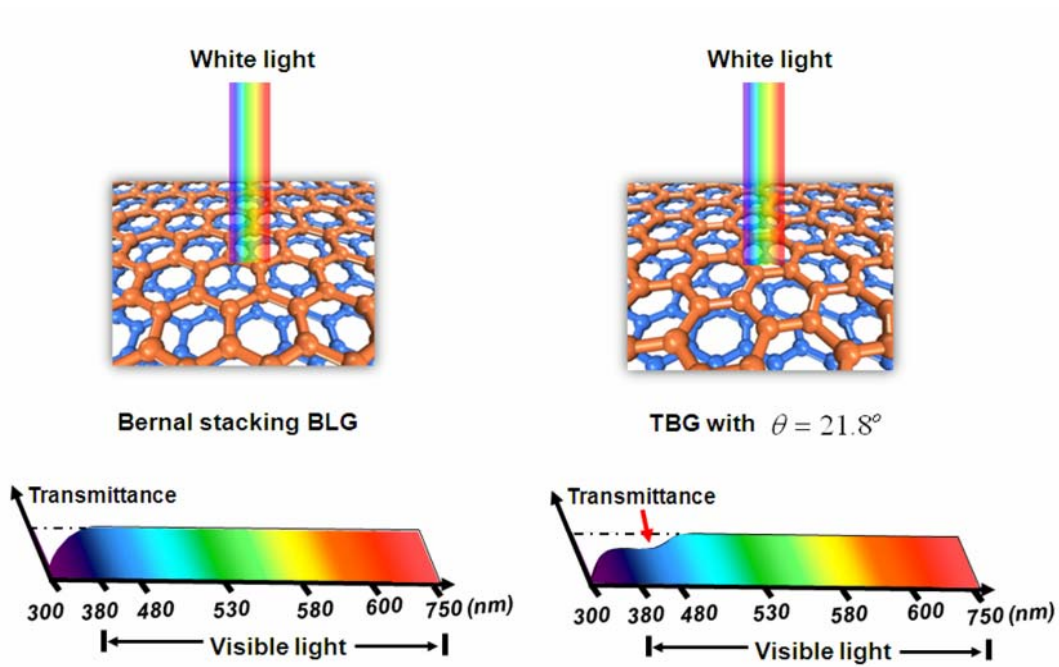

Figure 3 Left: Absorption of light by BLG with Bernal stacking. In the visible light range, there is a constant transmittance ~ 95.4% (absorption ~ 4.6%) due to the universal optical conductivity value. At the ultraviolet light range (300 nm-380 nm), the deviation from constant absorption is induced by the electronic transition near M point. Right: Absorption of light by the TBG sample with 21.8º orientation angle. Beyond the absorption at the ultraviolet light range, there is additional absorption peak in the visible light range (highlighted by a red arrow) which is induced by the electronic transition at M point with transition energy of ~ 2.77 eV.

Figure 4

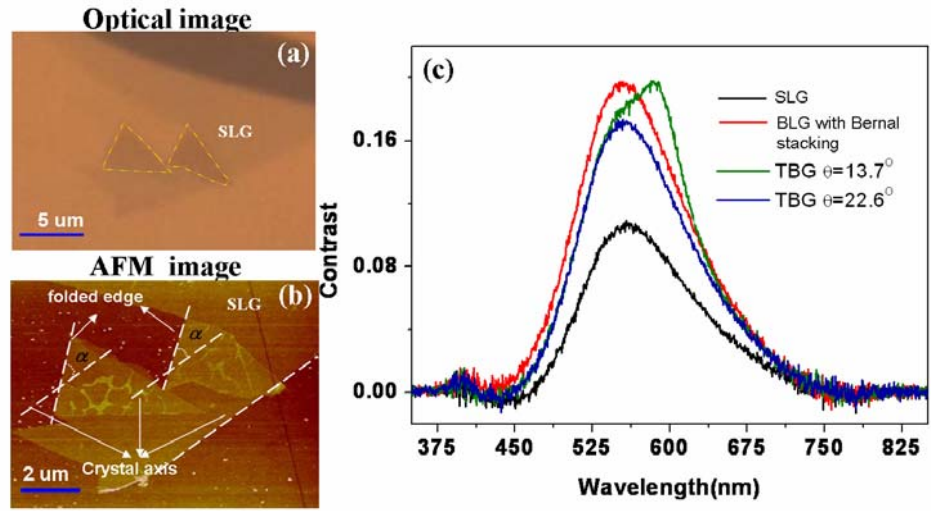

Figure 4. (a) Optical image of a piece of graphene sample on SiO$_2$/Si substrate that contains two folded parts which can be taken as two TBG samples. The boundaries of the TBG samples are demarcated by the yellow dashed lines. The thickness of SiO$_2$ is 272.0 ± 0.8 nm measured by ellipsometry. (b) AFM image of the TBG samples. The crystal axis (*i.e.* edge of the SLG) as well as folding line is shown by the white dashed line. The angle between an crystal axis and folding line is marked by $\alpha$, and according to the geometry analysis, the orientation angle $\theta$, which is the angle of the top layer rotates relative to the bottom layer can be determined as $\theta = 2\alpha$ or $\theta = 180° - 2\alpha$. The orientation angle $\theta = 22.6°$ for the left side TBG sample and $\theta = 13.7°$ for the right side TBG sample. (c) The contrast spectra of SLG, BLG with Bernal stacking and those two TBG samples.

Figure 5

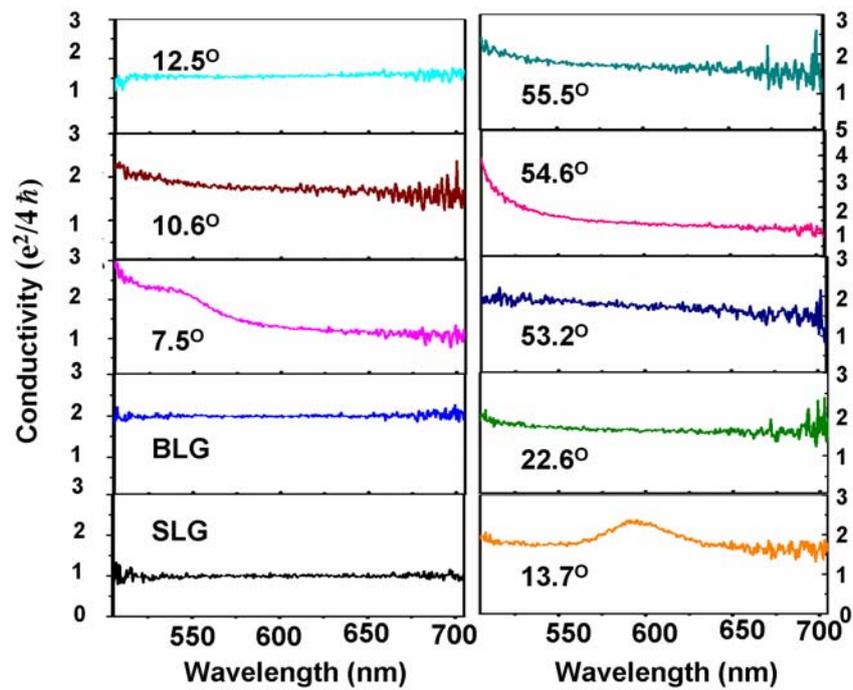

Figure 5. The optical conductivities of SLG, BLG with Bernal stacking, and TBG samples with orientation angle of $\theta$=7.5°, 10.6°, 12.5°, 13.7°, 22.6°, 53.2°, 54.6° and 55.5° which are obtained from their contrast spectra.

**REFERENCES AND NOTES**

1. Novoselov, K. S.; Geim, A. K.; Morozov, S. V.; Jiang, D.; Zhang, Y.; Dubonos, S. V.; Grigorieva, I. V.; Firsov, A. A. Electric Field Effect in Atomically Thin Carbon Films. *Science* **2004**, *306*, 666-669.

2. Berger, C.; Song, Z. M.; Li, X. B.; Wu, X. S.; Brown, N.; Naud, C.; Mayou, D.; Li, T. B.; Hass, J.; Marchenkov, A. N.; *et al.* Electronic Confinement and Coherence in Patterned Epitaxial Graphene. *Science* **2006**, *312*, 1191-1196.

3. Geim, A. K.; Novoselov, K. S. The Rise of Graphene. *Nat. Mater.* **2007**, *6*, 183-191.

4. Novoselov, K. S.; Geim, A. K.; Morozov, S. V.; Jiang, D.; Katsnelson, M. I.; Grigorieva, I. V.; Dubonos, S. V.; Firsov, A. A. Two-dimensional Gas of Massless Dirac Fermions in Graphene. *Nature* **2005**, *438*, 197-200.

5. Zhang, Y. B.; Tan, Y. W.; Stormer, H. L.; Kim, P. Experimental Observation of The Quantum Hall Effect and Berry's Phase in Graphene. *Nature* **2005**, *438*, 201-204.

6. Nomura, K.; MacDonald, A. H. Quantum Transport of Massless Dirac Fermions. *Phys. Rev. Lett.* **2007**, *98*, 076602/1-/4.

7. Bolotin, K. I.; Sikes, K. J.; Jiang, Z.; Klima, M.; Fudenberg, G.; Hone, J.; Kim, P.; Stormer, H. L. Ultrahigh Electron Mobility in Suspended Graphene. *Solid State Commun.* **2008**, *146*, 351-355.

8. Hwang, E. H.; Adam, S.; Das Sarma, S. Carrier Transport in Two-Dimensional Graphene Layers. *Phys. Rev. Lett.* **2007**, *98*, 186806/1-/4.

9. Falkovsky, L. A.; Varlamov, A. A. Space-Time Dispersion of Graphene Conductivity. *Eur. Phys. J. B* **2007**, *56*, 281-284.

10. Stauber, T.; Peres, N. M. R.; Geim, A. K. Optical Conductivity of Graphene in The Visible Region of The Spectrum. *Phys. Rev. B* **2008**, *78,* 085432/1-/8.

11. Nair, R. R.; Blake, P.; Grigorenko, A. N.; Novoselov, K. S.; Booth, T. J.; Stauber, T.; Peres, N. M. R.; Geim, A. K. Fine Structure Constant Defines Visual Transparency of Graphene. *Science* **2008**, *320*, 1308-1308.

12. Mak, K. F.; Sfeir, M. Y.; Wu, Y.; Lui, C. H.; Misewich, J. A.; Heinz, T. F. Measurement of The Optical Conductivity of Graphene. *Phys. Rev. Lett.* **2008**, *101,* 196405/1-/4.

13. Fei, Z.; Shi, Y.; Pu, L.; Gao, F.; Liu, Y.; Sheng, L.; Wang, B. G.; Zhang, R.; Zheng, Y. D. High-Energy Optical Conductivity of Graphene Determined By Reflection Contrast Spectroscopy. *Phys. Rev. B* **2008**, *78,* 201402/1-/4.

14. Min, H.; MacDonald, A. H. Origin of Universal Optical Conductivity and Optical Stacking Sequence Identification in Multilayer Graphene. *Phys. Rev. Lett.* **2009**, *103*,067402/1-/4.

15. Wang, F.; Zhang, Y. B.; Tian, C. S.; Girit, C.; Zettl, A.; Crommie, M.; Shen, Y. R. Gate-Variable Optical Transitions in Graphene. *Science* **2008**, *320,* 206-209.

16. Nicol, E. J.; Carbotte, J. P. Optical Conductivity of Bilayer Graphene With and Without an Asymmetry Gap. *Phys. Rev. B* **2008**, *77,*155409/1-/11.

17. Kresse, G.; Hafner, J. Ab-Initio Molecular-Dynamics For Liquid-Metals. *Phys. Rev. B* **1993**, *47*, 558-561.

18. Kresse, G.; Hafner, J. Ab-Initio Molecular-Dynamics For Open-shell Transition-Metals. *Phys. Rev. B* **1993**, *48*, 13115-13118.

19. Ni, Z. H.; Liu, L.; Wang, Y. Y.; Zheng, Z.; Li, L. J.; Yu, T.; Shen, Z. X. G-band Raman Double Resonance in Twisted Bilayer Graphene: Evidence of Band Splitting and Folding. *Phys. Rev. B* **2009**, *80*,125404/1-/5.

20. Ohta, T.; Bostwick, A.; McChesney, J. L.; Seyller, T.; Horn, K.; Rotenberg, E. Interlayer Interaction and Electronic Screening in Multilayer Graphene Investigated with Angle-resolved Photoemission Spectroscopy. *Phys. Rev. Lett.* **2007**, *98*, 206802/1-/4.

21. Gruneis, A.; Attaccalite, C.; Pichler, T.; Zabolotnyy, V.; Shiozawa, H.; Molodtsov, S. L.; Inosov, D.; Koitzsch, A.; Knupfer, M.; Schiessling, J.; *et al.* Electron-Electron Correlation in Graphite: a Combined Angle-Resolved Photoemission and First-Principles Study. *Phys. Rev. Lett.* **2008**, *100*, 037601/1-/4.

22. Trevisanutto, P. E.; Giorgetti, C.; Reining, L.; Ladisa, M.; Olevano, V. Ab Initio GW Many-Body Effects in Graphene. *Phys. Rev. Lett.* **2008**, *101*, 226405/1-/4.

23. McCann, E.; Fal'ko, V. I. Landau-Level Degeneracy and Quantum Hall Effect in a Graphite Bilayer. *Phys. Rev. Lett.* **2006**, *96*, 086804/1-/4.

24. Dos Santos, J.; Peres, N. M. R.; Castro, A. H. Graphene Bilayer With a Twist: Electronic Structure. *Phys. Rev. Lett.* **2007**, *99*, 256802/1-/4.

25. Hass, J.; Varchon, F.; Millan-Otoya, J. E.; Sprinkle, M.; Sharma, N.; De Heer, W. A.; Berger, C.; First, P. N.; Magaud, L.; Conrad, E. H. Why Multilayer Graphene on 4H-SiC(000$\bar{1}$) Behaves Like a Single Sheet of Graphene. *Phys. Rev. Lett.* **2008**, *100*, 125504/1-/4.

26. Yang, L.; Deslippe, J.; Park, C. H.; Cohen, M. L.; Louie, S. G. Excitonic Effects on The Optical Response of Graphene and Bilayer Graphene. *Phys. Rev. Lett.* **2009**, *103*, 186802/1-/4.

27. Kuzmenko, A. B.; van Heumen, E.; Carbone, F.; van der Marel, D. Universal Optical Conductance of Graphite. *Phys. Rev. Lett.* **2008**, *100*, 117401/1-/4.

28. Du, C. L.; You, Y. M.; Kasim, J.; Ni, Z. H.; Yu, T.; Wong, C. P.; Fan, H. M.; Shen, Z. X. Confocal White Light Reflection Imaging for Characterization of Metal Nanostructures. *Opt. Commun.* **2008**, *281*, 5360-5363.

29. Ni, Z. H.; Wang, Y. Y.; Yu, T.; You, Y. M.; Shen, Z. X. Reduction of Fermi Velocity in Folded Graphene Observed by Resonance Raman Spectroscopy. *Phys. Rev. B* **2008**, *77*, 235403/1-/5.

30. You, Y. M.; Ni, Z. H.; Yu, T.; Shen, Z. X. Edge Chirality Determination of Graphene by Raman Spectroscopy. *Appl. Phys. Lett.* **2008**, *93*, 163112/1-/3.

31. Ni, Z. H.; Wang, H. M.; Kasim, J.; Fan, H. M.; Yu, T.; Wu, Y. H.; Feng, Y. P.; Shen, Z. X. Graphene Thickness Determination Using Reflection and Contrast Spectroscopy. *Nano Lett.* **2007**, *7*, 2758-2763.

32. Casiraghi, C.; Hartschuh, A.; Lidorikis, E.; Qian, H.; Harutyunyan, H.; Gokus, T.; Novoselov, K. S.; Ferrari, A. C. Rayleigh Imaging of Graphene and Graphene Layers. *Nano Lett.* **2007**, *7*, 2711-2717.

33. Roddaro, S.; Pingue, P.; Piazza, V.; Pellegrini, V.; Beltram, F. The Optical Visibility of Graphene: Interference Colors of Ultrathin Graphite on $SiO_2$. *Nano Lett.* **2007**, *7*, 2707-2717.

34. Blake, P.; Hill, E. W.; Neto, A. H. C.; Novoselov, K. S.; Jiang, D.; Yang, R.; Booth, T. J.; Geim, A. K. Making Graphene Visible. *Appl. Phys. Lett.* **2007**, *91*, 063124/1-/3.

35. The conductivity error will be amplified when the contrast value is approach to 0. For contrast value C=0.0939±0.005, the calculated conductivity error is ~ 11.7%, and for contrast value C=0.0059±0.005, the calculated conductivity error is ~ 135.4%, which makes the calculated conductivity unreliable at contrast value approach 0.

36. We have obtained the optical conductivity of single to four layers graphene and they are almost the same after divided by the layer numbers, and are frequency dependent. This indicates such deviation from the predicted universal optical conductivity of graphene is due to the artifact of the system, most probably the NA effect. Therefore, normalization is carried for the experimental curves, such that the conductivity of single and few layers graphene are frequency independent, following the theoretical prediction.

37. Berger, C.; Song, Z. M.; Li, T. B.; Li, X. B.; Ogbazghi, A. Y.; Feng, R.; Dai, Z. T.; Marchenkov, A. N.; Conrad, E. H.; First, P. N.; *et al.* Ultrathin Epitaxial Graphite: 2D Electron Gas Properties and a Route Toward Graphene-Based Nanoelectronics. *J. Phys. Chem. B* **2004**, *108*, 19912-19916.

38. Chen, W.; Xu, H.; Liu, L.; Gao, X. Y.; Qi, D. C.; Peng, G. W.; Tan, S. C.; Feng, Y. P.; Loh, K. P.; Wee, A. T. S. Atomic Structure of The 6H-SiC(0001) Nanomesh. *Surf. Sci.* **2005**, *596*, 176-186.

39. Sutter, P. W.; Flege, J. I.; Sutter, E. A. Epitaxial Graphene on Ruthenium. *Nat. Mater.* **2008**, *7*, 406-411.

40. Reina, A.; Jia, X. T.; Ho, J.; Nezich, D.; Son, H. B.; Bulovic, V.; Dresselhaus, M. S.; Kong, J. Large Area, Few-Layer Graphene Films on Arbitrary Substrates by Chemical Vapor Deposition. *Nano Lett.* **2009**, *9*, 30-35.

41. Kim, K. S.; Zhao, Y.; Jang, H.; Lee, S. Y.; Kim, J. M.; Ahn, J. H.; Kim, P.; Choi, J. Y.; Hong, B. H. Large-Scale Pattern Growth of Graphene Films for Stretchable Transparent Electrodes. *Nature* **2009**, *457*, 706-710.

42. Li, X. S.; Cai, W. W.; An, J. H.; Kim, S.; Nah, J.; Yang, D. X.; Piner, R.; Velamakanni, A.; Jung, I.; Tutuc, E. *et al.* Large-Area Synthesis of High-Quality and Uniform Graphene Films on Copper Foils. *Science* **2009**, *324*, 1312-1314.

43. Varchon, F.; Mallet, P.; Magaud, L.; Veuillen, J. Y. Rotational Disorder in Few-Layer Graphene Films on 6H-SiC(000-1): a Scanning Tunneling Microscopy Study. *Phys. Rev. B* **2008**, *77*, 165415/1-/5.